\documentclass[reprint,           
               showpacs,            
               preprintnumbers,     
               aps,                 
               prd,          	    
               letterpaper,             
               groupaddress,      
               nofootinbib,         
               tightenlines,        
               floats,floatfix      
               ]{revtex4-1}

\usepackage{graphicx}
\usepackage{dcolumn}
\usepackage{bm}
\usepackage{amsmath}
\usepackage{epstopdf}
\usepackage{latexsym}

\begin{document}

\title{Non-relativistic approach for cosmological Scalar Field Dark Matter}

\author{L. Arturo Ure\~na-L\'opez}%
 \email{lurena@fisica.ugto.mx}
\affiliation{%
Departamento de F\'isica, DCI, Campus Le\'on, Universidad de
Guanajuato, 37150, Le\'on, Guanajuato, M\'exico.}

\date{\today}

\begin{abstract}
We derive non-relativistic equations of motion for the formation of cosmological structure in a Scalar Field Dark Matter (SFDM) model corresponding to a complex scalar field endowed with a quadratic scalar potential. Starting with the equations of motion written in the Newtonian gauge of scalar perturbations, we separate out the involved fields into relativistic and non-relativistic parts, and find the equations of motion for the latter that can be used to build up the full solution. One important assumption will be that the SFDM field is in the regime of fast oscillations, under which its behavior in the homogeneous regime is exactly that of cold dark matter. The resultant equations are quite similar to the Schr\"odinger-Poisson system of Newtonian boson stars plus relativistic leftovers, and they can be used to study the formation of cosmological structure in SFDM models, and others alike, to ultimately prove their viability as complete dark matter models.
\end{abstract}

\pacs{98.80.-k,95.35.+d,98.80.Jk}
\keywords{Cosmology, scalar fields, dark matter}
\maketitle

The nature of dark matter is one of the most puzzling topics in Cosmology today, and, not surprisingly, a large part of the specialized research is currently devoted to hound this elusive matter that makes out around $25\%$ of the total matter budget of the present Universe\cite{Ade:2013zuv,*nasa}. On the theoretical side, there are many models with proposals about particles and fields, all beyond our current understanding of particle physics, that are studied in detail to explain the bunch of cosmological observations we now have at hand\cite{Popolo:2012bna}.

A dark matter model that has captured the attention of different research groups is that comprised of a scalar field, which has been generically called Scalar Field Dark Matter (SFDM), in contrast to the standard Cold Dark Matter (CDM) model. This type of models has a very long tradition in the specialized literature, and their properties have been extensively studied under diverse circumstances, from the cosmological setting to the self-gravitational collapse of arbitrary configurations. The recent discovery of the Higgs boson\cite{Aad:2012tfa,*Chatrchyan:2012ufa}, and then the possible existence of fundamental scalars, has also renewed the interest in them. Some excellent reviews about the wide capabilities of SFDM can be found in\cite{Matos:2004rs,*Matos:2000ng,*Matos:2000ss,*Matos:1999et,Sahni:1999qe,*Peebles:2000yy,*Park:2012ru,Boehmer:2007um,*Harko:2011jy,Magana:2012xe,*Suarez:2013iw,Li:2013nqa,*Rindler-Daller:2013zxa,*AmaroSeoane:2010qx,*Briscese:2011ka}.

The purpose of this paper is to provide simple equations of motion for the formation of cosmological structure in SFDM models, and one key assumption for that is the explicit separation of the fields in relativistic and non-relativistic parts. As we shall show, if the relativistic part is properly taken off, the non-relativistic part of the fields obey equations of motion that are similar to the well known Schr\"odinger-Poisson (SP) system, which in turn can reduce the numerical complexity to follow the gravitational collapse of SFDM within an expanding cosmological setting.

To begin with, we discuss the general procedure to separate out the relativistic and non-relativistic parts in a field that obeys the Klein-Gordon (KG) equation of relativistic field theory. Common wisdom in textbooks (e.g.\cite{Ryder:1985wq}) suggests that the KG equation is a direct expression of the relativistic energy equation $E^2 = m^2 c^4 + p^2 c^2$, and then it only suffices to make the substitutions $E \to i\hbar \partial_t$, and $p \to -i\hbar \nabla$, to obtain the famous linear differential equation for the scalar field $\phi$: $-(\hbar^2/m^2 c^4) \ddot{\phi} = \phi - (\hbar^2/m^2c^2) \nabla^2 \phi$, where a dot means derivative with respect to the time $t$. Notice that we can read out from the KG equation the natural scales for time and distance in the evolution of the scalar field: these are the Compton length $L_C \equiv h/(mc)$, and the Compton time $T_C \equiv h/(m c^2) = L_C/c$. Likewise, if we take the non-relativistic expression for the energy, $E = p^2/2m$, then we obtain the equally famous Schr\"odinger equation for the wave function $\psi$: $i\hbar \dot{\psi} = -(\hbar^2/2m) \nabla^2 \psi$. Notice that this time, however, there are not preferred choices for the time and distance scales.

Even though they obey intrinsically different equations of motion, there exists an explicit relationship between the two scalar functions: $\phi = e^{-i 2\pi t/T_C} \psi$, under which the KG equation directly becomes:
\begin{equation}
  \label{eq:1}
  i\hbar \dot{\psi} - (\hbar^2/2m c^2) \ddot{\psi} = - (\hbar^2/2m) \nabla^2 \psi \, .
\end{equation}
Eq.~\eqref{eq:1} resembles the writing of the relativistic energy equation in the form: $(1+K/2mc^2)K = p^2/2m$, where $K = E - mc^2$; the differential equation is recovered under the identifications $K \to i\hbar \partial_t$ and $p \to i \hbar \nabla$. Eq.~\eqref{eq:1} is then an exact representation of the KG equation in terms of the wavefunction $\psi$; this fact has been used before, for instance, to find relativistic corrections of Bose-Einstein condensates\cite{Matos:2012qu,*Castellanos:2013kga}. 

The two extreme versions of the KG equation, namely the relativistic wave equation ($K^2/c^2 = p^2$), or the non-relativistic Schr\"odinger equation ($K= p^2/2m$), are easily obtained under the conditions $K \gg 2mc^2$, or $K \ll 2mc^2$, respectively. In terms of the differential operators, this is equivalent to say that the time derivative of the wave function $\psi$ is of the same or smaller order when compared to its spatial derivative. For example, if $\nabla \psi = \mathcal{O}(\epsilon)$ and $\partial_t \psi = \mathcal{O}(\epsilon^2)$, where $\epsilon$ is a small parameter, the Schr\"odinger equation is recovered if we keep in Eq.~\eqref{eq:1} only terms of the order $\epsilon^2$ in the differential operators, and drop out $\partial^2_t \psi = \mathcal{O}(\epsilon^4)$ for being of higher order in  $\epsilon$.

The overall argument above helps us to set up the philosophy in this work. By taking advantage of the explicit relationship between the relativistic fields and their non-relativistic counterparts, we calculate the equations of motion that are obeyed by the non-relativistic components together with the special features that are left behind in them by the field transformations. The non-relativistic equations can then be used to build up the full solution in most cases (except, of course, for the extremely relativistic ones), with less numerical effort. This kind of procedure has been used successfully before in the case of boson stars\cite{Seidel:1990jh,*Guzman:2004wj,*UrenaLopez:2001tw,*Ruffini:1969qy}, and we are about to show below its application for the cosmological setting of SFDM.

We shall consider the following scalar perturbations, in the Newtonian gauge, for a flat Friedmann-Robertson-Walker metric (more details can be found in\cite{Weinberg:2008zzc,*Dodelson:2003ft}):
\begin{equation}
  \label{eq:8}
  ds^2 = - (1+2\Psi) dt^2 + a^2(t) (1-2\Phi) \delta_{ij}dx^i dx^j \, ,
\end{equation}
where $a(t)$ is the scale factor of the Universe (we will use units in which $c=1=\hbar$ hereafter). The gravitational potentials $\Psi$ and $\Phi$ are small enough to be considered metric perturbations, $|\Psi|, \, |\Phi| \ll 1$; in fact, the homogeneous and isotropic metric is obtained in the limit $\Psi =0= \Phi$. It is known that anisotropic stress in the matter sources causes differences in the evolution of the gravitational potentials, but this is only relevant for the very early Universe. Then, a good approximation for the gravitational potentials is that $\Phi = \Psi$ for most of the evolution of the Universe, specially for the era of structure formation which is our main interest here\cite{Weinberg:2008zzc}. We will anyway write all formulas below in full generality, and only enforce the aforementioned condition later on for the main physical results.

The equations of motion are provided by Einstein's equations: $G_{\mu \nu} = 8\pi G \, \left( T^{(\Lambda)}_{\mu \nu} + T^{(\phi)}_{\mu \nu} \right)$, where $G_{\mu \nu}$ is the Einstein tensor, and $G$ is Newton's gravitational constant. As dark energy we adopt a cosmological constant $\Lambda$, with $T^{(\Lambda)}_{\mu \nu} = (\Lambda/8\pi G) g_{\mu \nu}$, and as SFDM we take a complex scalar field $\phi$ endowed with a quadratic potential of the form $V(\phi) = m^2 |\phi|^2/2$\cite{UrenaLopez:2000gm,Li:2013nqa}, with $m$ being the mass parameter of the scalar field. Its corresponding energy-momentum tensor is:
\begin{equation}
  \label{eq:14}
  T^{(\phi)}_{\mu \nu} = \frac{1}{2} (\partial_\mu \phi \partial_\nu \phi^\ast + \partial_\mu \phi^\ast \partial_\nu \phi) - \frac{1}{2} g_{\mu \nu} (\partial^\sigma \phi \partial_\sigma \phi^\ast + m^2 |\phi|^2) \, .
\end{equation}

We start by writing the equation of motion for the scalar field that arises from the conservation of the energy-momentum tensor, ${T^{(\phi)\mu \nu}}_{; \nu} =0$; using the metric~\eqref{eq:8} and Eq.~\eqref{eq:14}, we find that\cite{Matos:2009rw,Weinberg:2008zzc,Li:2013nqa}:
\begin{widetext}
\begin{equation}
  \label{eq:2}
  (1-2\Psi) \ddot{\phi} + (3H - \dot{\Psi} - 3 \dot{\Phi} - 6H \Psi) \dot{\phi} - a^{-2} (1+2\Phi) \nabla^2 \phi - a^{-2} \nabla (\Psi - \Phi) \cdot \nabla \phi + m^2 \phi = 0 \, .
\end{equation}
\end{widetext}

On the other hand, the time-time component of the Einstein equations results in the Poisson-like equation of motion for the gravitational potentials:
\begin{equation}
  \label{eq:6}
  \nabla^2 \Phi - 3 H a^2 ( \dot{\Phi} + H \Psi ) = 4\pi G a^2 \, \delta \rho \, .
\end{equation}
On the rhs we have the energy overdensity $\delta \rho = \rho_T - \bar{\rho}_T$, where $\rho_T = \rho_\phi + \rho_\Lambda$ is the total energy density, and $\bar{\rho}_T \equiv  3 H^2/(8\pi G)$ is its homogeneous counterpart, with $H \equiv \dot{a}/a$ the Hubble parameter. (Barred variables will denote homogeneous quantities.) The energy density of the cosmological constant is homogeneous, and then the only energy overdensity is that of SFDM, i.e. $\delta \rho = \rho_\phi - \bar{\rho}_\phi$. The energy density of the scalar field reads
\begin{equation}
  \label{eq:13}
  \rho_\phi = -{T^{(\phi)0}}_0 = \frac{(1-2\Psi)}{2} |\dot{\phi}|^2 + \frac{(1+2\Phi)}{2a^2} |\nabla \phi|^2 + \frac{1}{2} m^2 |\phi|^2 \, ,
\end{equation}
and then $\bar{\rho}_\phi = (1/2) ( |\dot{\bar{\phi}}|^2 + m^2 |\bar{\phi}|^2)$.

Next, we set the changes $\phi = a^{-3/2} e^{-imt} \psi$ and $\Phi=\Psi$. After lengthy but otherwise straightforward calculations, Eq.~\eqref{eq:2} and~\eqref{eq:13} become:
\begin{widetext}
  \begin{subequations}
    \label{eq:3}
    \begin{eqnarray}
      \label{eq:3a}
      \frac{(1-2\Psi)}{2} (\ddot{\psi} - 2im \dot{\psi}) -2 H \Psi  (\dot{\psi} - im \psi) = \frac{(1+2\Psi)}{2a^2} \nabla^2 \psi - \left[ \left( m^2 + 2 H^2 \right)  \Psi  - \frac{3}{4} \left( \dot{H} + \frac{3}{2} H^2 \right) \left( 1 - 2\Psi \right) \right] \psi \, , \\
      \label{eq:3b}
      a^3 \rho_\phi = \frac{(1-2\Psi)}{2} \left[ |\dot{\psi}|^2 -3 H \, {\rm Re}(\dot{\psi} \psi^*) + 2 m \, {\rm Im}(\dot{\psi} \psi^*) \right] +  \frac{ (1+2\Psi)}{2a^2} |\nabla \psi|^2 + m^2 |\psi|^2 \left[ 1 +  \frac{9}{8} \frac{H^2}{m^2} - \Psi \left( 1 + \frac{9}{4} \frac{H^2}{m^2} \right) \right] \, .
  \end{eqnarray}
\end{subequations}
We now apply the non-relativistic approximation of the gravitational potential, which will be considered to obey the order relation: $\Psi = \mathcal{O}(\epsilon^2)$, and then will be neglected wherever it appears alongside with terms of lower order in $\epsilon$; another simplifying step is to consider the decomposition $\Psi (t, \mathbf{x}) = \Psi_0(t,\mathbf{x})/a(t)$, so that $\dot{\Psi} = - H \Psi + \dot{\Psi}_0/a$. It is also convenient to work with the following dimensionless variables: $\sqrt{4\pi G} \psi \to \psi$, $m t \to t$, and $m \mathbf{x} \to \mathbf{x}$. Eqs.~\eqref{eq:3} then become
  \begin{subequations}
    \label{eq:3v}
    \begin{eqnarray}
      \label{eq:3va}
      \frac{1}{2a^2} \nabla^2 \psi - \frac{1}{2} \ddot{\psi} + i \dot{\psi} &=&   \frac{1}{a} \left[ 1 + 2 \frac{H^2}{m^2} - 2 i \frac{H}{m} + \frac{9}{4} \frac{H^2}{m^2} \left( 1 + \frac{2}{3} \frac{\dot{H}}{H^2} \right) \right] \Psi_0 \psi - \frac{9}{8} \frac{H^2}{m^2} \left( 1 + \frac{2}{3} \frac{\dot{H}}{H^2} \right) \psi \, , \\
      \label{eq:3vb}
      \nabla^2 \Psi_0 - 3\frac{H}{m} a^2 \dot{\Psi}_0 &=& \frac{1}{2} |\dot{\psi}|^2 - \frac{3}{2} \frac{H}{m} \, {\rm Re}(\dot{\psi} \psi^*) + {\rm Im}(\dot{\psi} \psi^*) +  \frac{ |\nabla \psi|^2}{2a^2} + \left( |\psi|^2 - |\bar{\psi}|^2 \right) \left( 1 + \frac{9}{8} \frac{H^2}{m^2} \right) \, ,
  \end{eqnarray}
\end{subequations}
where $\bar{\rho}_\phi$ was calculated under the assumption that $\bar{\phi} = a^{-3/2} e^{-imt} \bar{\psi}$, with $\bar{\psi} = \mathrm{const}$.
\end{widetext}

Eqs.~\eqref{eq:3v} are the equations of motion for SFDM that result once we separate out the relativistic oscillations of the scalar field, and consider that the gravitational potential $\Psi$ has always a small amplitude. We can apply one further simplification for the Hubble parameter $H$ and its time derivative $\dot{H}$. which appear as ubiquitous companions of the mass term $m$ everywhere in Eqs.~\eqref{eq:3v}, the so called limit of fast oscillations $H/m \ll 1$, under which the scalar field $\phi$ behaves exactly like CDM in the homogeneous regime. According to previous studies, the regime of fast oscillations must be present already before the time of radiation-matter equality for a good consistency with cosmological observations\cite{Matos:2004rs,*Matos:2000ng,*Matos:1999et,Sahni:1999qe,*Peebles:2000yy,*Park:2012ru,Boehmer:2007um,*Harko:2011jy,Magana:2012xe,*Suarez:2013iw,Li:2013nqa}. 

The same applies for the equation of motion~\eqref{eq:3vb}, which has the form of an inhomogeneous heat equation for the gravitational potential $\Psi$. The companion coefficient of $\dot{\Psi}$ (which would play the role of a thermal diffusivity) is a growing function of time, but the regime of fast oscillations makes it anyway negligible for the relevant period of structure formation.

As for $\dot{H}$, the equations of motion of a homogeneous and isotropic Universe\cite{Weinberg:2008zzc} show that
\begin{eqnarray}
  \label{eq:5}
  \frac{9}{8} \frac{H^2}{m^2} \left( 1+ \frac{2}{3} \frac{\dot{H}}{H^2} \right) = - \frac{3\pi G}{m^2} \bar{p}_T 
  = - \frac{3\pi G}{m^2} \bar{w} \bar{\rho}_T \nonumber \\
  = - \frac{9}{8} \frac{H^2}{m^2} \times \left\{ 
    \begin{array}{lr}
      1/3 & {\rm RD} \\
      0 & {\rm MD} \\
      -1 & \Lambda {\rm D}
    \end{array}
  \right. \, ,
\end{eqnarray}
where $\bar{p}_T = \bar{p}_T(t)$ ($\bar{w}$) is the total (homogeneous and isotropic) pressure (equation of state) provided by all matter fluids in the cosmos, and the labels stand for each one of the main stages in the evolution of the Universe: Radiation Domination (RD), Matter Domination (MD), and $\Lambda$ Domination  $\Lambda$D. Thus, the correction induced by $\dot{H}$ is always directly proportional to the ratio $H^2/m^2$, and then it also becomes practically negligible in the regime of fast oscillations. 

The surviving, leading order terms in Eqs.~\eqref{eq:3v} are:
\begin{subequations}
  \label{eq:4}
    \begin{eqnarray}
      \label{eq:4a}
      \frac{1}{2a^2} \nabla^2 \psi &-& \frac{1}{2} \ddot{\psi} + i \dot{\psi} =  \frac{1}{a} \Psi_0 \psi \, , \\
      \label{eq:4b}
      \nabla^2 \Psi_0 &=& \frac{1}{2} |\dot{\psi}|^2 + \, {\rm Im}(\dot{\psi} \psi^*) +  \frac{ |\nabla \psi|^2}{2a^2} + |\psi|^2 - |\bar{\psi}|^2 \, .
    \end{eqnarray}
  \end{subequations}
Eqs.~\eqref{eq:4} are the main results in this paper, and represent the equations of motion for the non-relativistic formation of structure in the SFDM model. They look more involved than the usual SP system that appears under the direct application of Newtonian cosmology for sub-Hubble scales, see for instance\cite{Hu:2000ke}. 

To have a closed system of equations, we must add the equation of motion corresponding to the homogeneous and isotropic expansion of the Universe. In order to be consistent with our non-relativistic approach, we must write the Friedmann equation as
\begin{equation}
  \label{eq:9}
  \frac{H^2}{m^2} = \frac{2}{3} \frac{|\bar{\psi}|^2}{a^3} +  \tilde{\Lambda} \quad \Rightarrow \quad \dot{a} = \sqrt{\frac{2}{3}} a^{-1/2} \left( |\bar{\psi}|^2 +  \tilde{\Lambda} a^3 \right)^{1/2} \, .
\end{equation}
By taking into account that the cosmological constant provides a constant energy density, we have defined $\tilde{\Lambda} = (\Lambda/2m^2) = (3H^2_0/2m^2) \Omega_{\Lambda 0}$, where $H_0 = 67 \, {\rm km \, s^{-1}Mpc^{-1}}$ and $\Omega_{\Lambda,0} = 0.68$, are the present values of the Hubble constant and of the density parameter of the cosmological constant\cite{nasa}, respectively. As discussed above, $\bar{\psi} = {\rm const.}$, and then the scale factor $a(t)$ evolves exactly as in the standard $\Lambda$CDM model.

We will not attempt here to solve the equations of motion~\eqref{eq:4} to find their main features, but take an indirect approach and compare them with their Newtonian counterpart. As explained before, we take the following order estimations for the wavefunction and its spacetime derivatives: $\psi =  \mathcal{O}(\epsilon^2)$, $\nabla^2 \psi = \mathcal{O}(\epsilon^4)$, $\dot{\psi} = \mathcal{O}(\epsilon^4)$, and $\ddot{\psi} = \mathcal{O}(\epsilon^6)$, where $\epsilon$ is a small parameter. After this, the surviving terms in Eqs.~\eqref{eq:4} are
\begin{subequations}
  \label{eq:4v}
\begin{eqnarray}
  \label{eq:4va}
  i\dot{\psi} &=& - \frac{1}{2a^2} \nabla^2 \psi + \frac{1}{a} \Psi_0 \psi \, , \\
  \label{eq:4vb}
  \nabla^2 \Psi_0 &=& |\psi|^2 - |\bar{\psi}|^2 \, ,
\end{eqnarray}
\end{subequations}
which is the SP system that is widely used in the specialized literature.

As an example, let us consider the scaling properties of the equations of motion, much in the form that is commonly used in Newtonian boson stars\cite{Guzman:2004wj}. Eqs.~\eqref{eq:9} and ~\eqref{eq:4v} are invariant under the following scale transformation:
\begin{equation}
  \label{eq:11}
  \{ t, \mathbf{x}, \psi, \Psi_0, \tilde{\Lambda} \} \, \to \, \{ t/\lambda^2, \mathbf{x}/\lambda, \lambda^2 \psi, \lambda^2 \Psi_0, \lambda^4 \tilde{\Lambda} \} \, ,
\end{equation}
for any arbitrary parameter $\lambda$. This scaling invariance can ease the numerical effort, mainly because the fields themselves can have a small amplitude whereas their derivatives can be very large. An appropriate small value of $\lambda$ can help us to cancel such differences and to keep all quantities of order unity in numerical simulations\cite{Guzman:2004wj,oai:arXiv.org:0806.0232}. 

One natural possibility for the scaling of the equations of motion in the cosmological setting is $\lambda = \tilde{\Lambda}^{1/4} \ll 1$, so that $ \tilde{\Lambda}$ becomes the scale of reference for all physical quantities in the numerical simulation. For instance, the natural distance and time scales for the evolution of the fields would be $L = L_C/\lambda$ and  $T = T_C/\lambda^2$. Explicitly,
\begin{equation}
  \label{eq:12}
  L = \frac{1}{(3\Omega_{\Lambda,0})^{1/4}} \sqrt{\frac{m}{H_0}} L_C \, , \;  T = \frac{1}{(3\Omega_{\Lambda,0})^{1/2}} \frac{m}{H_0} T_C \, ,
\end{equation}
where $L_C$ and $T_C$ are the Compton length and time defined before. For the preferred case of SFDM with an ultralight scalar field, we find:
\begin{equation}
  \label{eq:15}
  \frac{L}{{\rm kpc}} \simeq 45 \sqrt{\frac{10^{-22}{\rm eV}}{m}}  \, , \quad \frac{T}{{\rm Myr}} \simeq 0.15 \sqrt{\frac{10^{-22}{\rm eV}}{m}} \, ,
\end{equation}
which are of the expected order of magnitude for a cosmological evolution if $m$ is small enough.

Eq.~\eqref{eq:11} shows that the SP system is a free-scale system, and then the complete set of gravitational solutions is a one-parameter family. This means that if a self-gravitating configuration is allowed to accrete matter, it will migrate to another equilibrium configuration which is denser and more compact, and for that it suffices to consider a scaled system with a larger $\lambda$\cite{Guzman:2004wj}. In principle, the migration process could continue endlessly up to the point that $\psi \to \infty$ and $\mathbf{x} \to 0$ for $\lambda \to \infty$. This possibility seems to have been observed in cosmological simulations in which is argued that SFDM leads to cusp density profiles of collapsed objects\cite{oai:arXiv.org:0806.0232,*Woo:2002tm,*Madarassy:2012bm}. It appears counterintuitive at first sight because of the wave nature of the Schr\"odinger equation, but is in fact a side effect of Eq.~\eqref{eq:11}. 

The above arguments are spoiled by the extra spacetime derivatives of the wavefunction that appear in Eqs.~\eqref{eq:4} and that were neglected in order to get the Newtonian equations~\eqref{eq:4v}. The scaling transformation~\eqref{eq:11} actually shows us that the extra terms can be neglected as long as $\lambda \ll 1$, but they must be taken into account once this premise is no longer satisfied. For our example above, this means that the SP system~\eqref{eq:4v} can only be reliable for gravitational systems with a size of the order or larger than $45$ kpc, as smaller systems would require the assistance of the non-Newtonian extra terms.

Some final comments are in turn. We have worked out in detail the equations of motion that are appropriate to study the formation of structure in SFDM models, and found that there must be extra terms in consideration apart form the standard structure of the SP system. The latter is indeed a good approximation for systems that evolve on large scales and times, and its scale invariance helps to ease the numerical efforts in simulations. But, for smaller systems the wave nature of the scalar field demands the more general set of equations in~\eqref{eq:4}, which is anyway more tractable than the original Einstein-Klein-Gordon equations of motion.

A more complete model of SFDM seems to require the presence of a quartic self-interaction in the scalar potential in order to avoid any disturbing behavior of the matter scalar field in an early RD era; such an interaction would be needed also to understand the condensation properties of the scalar field more properly\cite{UrenaLopez:2008zh,*Lundgren:2010sp,*Berges:2014xea,Matos:2012qu,*Castellanos:2013kga,Briscese:2011ka,*AmaroSeoane:2010qx,*Li:2013nqa,*Boehmer:2007um}. However, we have not included a quartic term because its presence would be non-negligible only well before the beginning of structure formation\cite{Li:2013nqa,Matos:2000ng,Sahni:1999qe,Boehmer:2007um,Magana:2012xe}.

On a quite different topic, we want to mention a line of research that is pursuing the study of structure formation by translating fluid equations of motion into the SP system: one of the aims is to have more friendly variables from the numerical perspective\cite{Davies:1996kp,*Coles:2002sj,*Schaller:2013zda,*Thomson:2011tg}. Those methods have not been explored exhaustively yet, even though they provide an alternative interpretation of the distribution function of $N$-body simulations in terms of a wave function. 

More important is that the equations of motion solved in those studies are quite similar to that of SFDM. Thus, we can anticipate that some of the intrinsic characteristics of the formation of structure under the SFDM hypothesis may have been already found in\cite{Davies:1996kp,*Coles:2002sj,*Schaller:2013zda,*Thomson:2011tg}, and also in the studies of the gravitational collapse of Newtonian boson stars\cite{Matos:2007zza,*Bernal:2009zy,*UrenaLopez:2010ur,Guzman:2003kt,*Guzman:2004wj,*Guzman:2005dw,*Guzman:2013rua,*Chavanis:2011zi,*Chavanis:2011zm,*Lora:2011yc,*RindlerDaller:2011kx}, all of them giving an indirect confirmation that SFDM works as well as CDM in the arena of structure formation in the Universe. However, the full characteristics of SFDM structure can only arise from the correct non-relativistic equations of motion~\eqref{eq:4} found above. This requires an extensive numerical study that goes beyond the purposes of the present paper and that will be reported elsewhere.

\begin{acknowledgments}
I am grateful with Tonatiuh Matos, and Francisco S. Guzm\'an for useful comments. This work was partially supported by PROMEP, DAIP, by
CONACyT M\'exico under grant 167335, the Fundaci\'on Marcos Moshinksy, and the Instituto Avanzado de Cosmolog\'ia (IAC) collaboration.
\end{acknowledgments}

\bibliography{refs}

\end{document}